\documentclass{aa}  
\usepackage{graphicx}
\usepackage{txfonts}
\usepackage{color}
\usepackage{longtable}
\begin{document}

   \title{Detection of Yarkovsky acceleration in the context of precovery observations and the future Gaia catalogue}


 \offprints{J.Desmars, desmars@imcce.fr}
   \author{J. Desmars
          \inst{1,2}
          }

   \institute{Observat\'{o}rio Nacional, Rua Jos\'e Cristino 77, S\~{a}o Cristov\~{a}o, Rio de Janeiro CEP 20.921-400, Brazil\\
              \email{desmars@on.br}
\and
Institut de M\'ecanique C\'eleste et de Calcul des \'Eph\'em\'erides
                - Observatoire de Paris, UMR 8028 CNRS, 77 avenue Denfert-Rochereau,
                75014 Paris, France. 
              \email{desmars@imcce.fr}             }

   \date{}

 
  \abstract
   {The Yarkovsky effect is a weak non-gravitational force leading to a small variation of the semi-major axis of an asteroid. Using radar measurements and astrometric observations, it is possible to measure a drift in semi-major axis through orbit determination. }
   {This paper aims to detect a reliable drift in semi-major axis of near-Earth asteroids (NEAs) from ground-based observations and to investigate the impact of precovery observations and the future Gaia catalogue in the detection of a secular drift in semi-major axis.}
   {We have developed a precise dynamical model of an asteroid's motion taking the Yarkovsky acceleration into account and allowing the fitting of the drift in semi-major axis. Using statistical methods, we investigate the quality and the robustness of the detection.}
   {By filtering spurious detections with an estimated maximum drift depending on the asteroid's size, we found 46 NEAs with a reliable drift in semi-major axis in good agreement with the previous studies. The measure of the drift leads to a better orbit determination and  constrains  some physical parameters of these objects. Our results are in good agreement with the $1/D$ dependence of the drift and with the expected ratio of prograde and retrograde NEAs. We show that the uncertainty of the drift mainly depends on the length of orbital arc and in this way we highlight the importance of the precovery observations and data mining in the detection of consistent drift. Finally, we discuss the impact of Gaia catalogue in the determination of drift in semi-major axis.}
   {}


   \maketitle
%

\section{Introduction}
Highlighted in the early 20\textsuperscript{th} century, the Yarkovsky effect is a non-gravitational force due to an anisotropic emission of thermal radiation. It mainly depends on the size, the density, the spin, and the thermal characteristics of the asteroid. The effect has a diurnal component related to the asteroid rotation, which is generally dominant, and a seasonal component related to the orbital motion \citep{Vok2000}. 

The Yarkovsky effect leads to an acceleration on the asteroid that modifies its semi-major axis as a secular drift. It has important consequences for the long-term evolution of asteroids. In particular, the Yarkovsky effect could explain the transportation of asteroids from the
main belt region to the inner solar system \citep{Bottke2006}. It also becomes necessary for accurate orbit determination or prediction of trajectory of near-Earth asteroids (NEA), in particular, in the context of close approach or impact predictions \citep{Farnocchia2013b,Farnocchia2014,Chesley2014}. 

The first detection of Yarkovsky acceleration has been performed thanks to radar measurements of the asteroid (6489) Golevka \citep{Chesley2003}. \cite{Vok2008} measured the Yarkovsky effect using only astrometric observations of the asteroid (152563) 1992BF. 

More recently, \cite{Nugent2012} detected a measurable secular drift for 54 NEAs and \cite{Farnocchia2013} for 21 NEAs using more restrictive criteria. \cite{Nugent2012} used the Orbfit package\footnote{http://adams.dm.unipi.it/orbfit/} for orbit determination and \cite{Farnocchia2013} used JPL's Comet and Asteroid Orbit Determination Package and cross-checked their results with Orbfit.

In this paper, we perform an independent study of the detection of the drift in semi-major axis of NEAs. We develop our own dynamical model allowing the orbit determination and the measurement of a secular drift in semi-major axis (Sect.~\ref{S:orbit}). We aim to detect a secular drift for the 10953 NEAs\footnote{As of 12 May 2014.} (1629 numbered and 9324 unnumbered) from ground-based observations. We first verify the robustness and  quality of the detection with statistical methods (Sects.~\ref{Ss:accuracy}~\&~\ref{Ss:stability}) and finally, by introducing filters to avoid unreliable detections, we detect a consistent measurement for 46 NEAs (Sect.~\ref{S:results}). As a long period of observations appears necessary for an accurate detection, we analyse how data mining through precovery observations can help to refine the measurement of a secular drift (Sect.~\ref{S:datamining}). Finally, we highlight the expected improvement with the Gaia mission for the detection of Yarkovsky acceleration (Sect.~\ref{S:gaia}).

\section{Orbit determination}\label{S:orbit}
To perform the orbit determination and to detect the drift, we build a dynamical model of an asteroid's motion  taking  the main perturbations into account (Sect.~\ref{Ss:model}), and we add the Yarkovsky effect as a transverse force depending on the orbital elements and on a drift in semi-major axis (Sect.~\ref{Ss:yarko}). The observations and their treatment are presented in Sect.~\ref{Ss:obs}. Finally, the fitting process, used to determine the state-vector of the asteroid (i.e. the position and the velocity of the asteroid at a specific date) and the value of the drift $\dot{a}$, is presented in Sect.~\ref{Ss:fit}.

\subsection{Dynamical model}\label{Ss:model}
To measure a drift in semi-major axis of asteroids, we developed a numerical integration of the motion of an asteroid (NIMA, thereafter). The equations of the motion are numerically integrated with a Gauss-Radau method \citep{Everhart1985}. The dynamical model includes:
\begin{itemize}
\item the perturbations of the planets, Pluto and the Moon, whose positions are given by JPL Ephemeris DE421 \citep{DE421};
\item the perturbations of the ten main asteroids whose positions have been preliminary computed by NIMA and turned into Chebychev ephemeris to make quick computation. The masses of the main asteroids are provided in Table~\ref{T:mainast};  
\item the corrections of relativistic effects of Sun, Jupiter, and the Earth as described in \cite{Moyer2000};  
\item the Yarkovsky effect (see Sect.~\ref{Ss:yarko}). 
\end{itemize}

\begin{table}[h!]
\begin{center}
\caption{Masses of the ten biggest asteroids from \cite{DE421} used in NIMA.}\label{T:mainast}
\begin{tabular}{lr}
\hline
\hline
asteroid & mass ($\times 10^{-12} M_\odot)$\\
\hline
    (1) Ceres           &       468.517 \\
    (4) Vesta           &       132.844 \\
    (2) Pallas          &       100.985 \\
   (10) Hygiea          &        40.418 \\
  (704) Interamnia      &        18.566 \\
   (16) Psyche          &        16.826 \\
   (15) Eunomia         &        12.342 \\
  (511) Davida          &        12.342 \\
    (3) Juno            &        11.574 \\
   (52) Europa          &        10.203 \\
 \hline
\end{tabular}
\end{center}
\end{table}

All solar system objects in the dynamical model are considered  point masses. The equations of variation are also numerically integrated with the equations of motion  to fit the model to observations by a Levenberg-Marquardt algorithm (see Sect.~\ref{Ss:fit}). To be consistent with the planetary ephemeris, we use the masses of the asteroids provided by \cite{DE421} and given in Table~\ref{T:mainast}. The masses of some big asteroids are not well determined. For example, we noticed that the mass of (704) Interamnia is $4.751$km$^3$/s$^2$ in \cite{Farnocchia2013} whereas the mass is $2.464$km$^3$/s$^2$ in \cite{DE421}. Moreover, we compared our results for the 46 selected NEAs (see Sect.~\ref{Ss:selectedNEAs}) by including the 25 biggest asteroids from \cite{DE421} as well as  the 10 biggest asteroids, and we noticed only minor differences. Consequently, and also for time computation, we choose the ten biggest asteroids corresponding to a mass greater than $10^{-11} M_\odot$.

\subsection{Yarkovsky acceleration modelling}\label{Ss:yarko}
We consider the Yarkovsky effect  a comet-like model as presented in \cite{Marsden1973}, and we use the effect for non-gravitational perturbation. The effect is considered as a transverse force: 
\begin{equation}
\mathbf{{a}_Y}=A_2g(r)\mathbf{t}
\end{equation}
where $\mathbf{t}$ is the normalised transverse vector $(\mathbf{r}\times\mathbf{v})\times\mathbf{r}$ with $\mathbf{r}$ and $\mathbf{v}$   the heliocentric position and the velocity of the asteroid, respectively.

The expression $g(r)$ is an $r$-function developed in \cite{Marsden1973} and $A_2$ is a parameter associated with this force. As in \cite{Farnocchia2013}, we used a simple expression with $g(r)=(r_0/r)^2$ where $r_0=1~au$ is a normalising parameter.

The Gauss equations give the variation of the semi-major axis related to the transverse acceleration
\begin{equation*}
\frac{da}{dt} = \dot{a} = \frac{2a\sqrt{1-e^2}A_2 r_0^2}{nr^3}
\end{equation*}
where $a$ is the semi-major axis, $e$ the eccentricity, $n$ the mean motion, and $r$ is the heliocentric distance of the asteroid.

Consequently, by averaging the previous relation on an orbital period, the mean variation of $\dot{a}$ is
\begin{eqnarray*}
<\dot{a}>  & = & \frac{2A_2 r_0^2}{na^2(1-e^2)}.
\end{eqnarray*}

Finally, the Yarkovsky acceleration is written as a transverse force only depending on the orbital elements and the drift in semi-major axis (now written as $\dot{a}$):
\begin{equation}
\mathbf{{a}_Y}=\frac{n}{2}\frac{a^2(1-e^2)}{r^2}\left(\dot{a}\right)\mathbf{t}. 
\end{equation}

This expression is equivalent to those given in \cite{Chesley2008} and \cite{Mouret2011}, and is consistent with models in \cite{Farnocchia2013} where they estimated the parameter $A_2$. We can estimate either $A_2$ parameter or the mean drift in semi-major axis $\dot{a}$. Typical values for the drift are $10^{-3}$ to $10^{-4}$au/My \citep{Vok2000}, corresponding to a range of $10^{-13}$ to $10^{-15}$ au/d$^2$ for $A_2$ . Thereafter we use $10^{-4}$au/My as a unit for the drift in semi-major axis $\dot{a}$ and its uncertainty $\sigma_{\dot{a}}$, and $10^{-15}$au/d$^2$ for $A_2$ parameter and its uncertainty.

\subsection{Observations and treatement}\label{Ss:obs}
The asteroid observations come from the Minor Planet Center database\footnote{\url{http://www.minorplanetcenter.net}} as of 12 May 2014. Observations are also available on the AstDyS database\footnote{\url{http://hamilton.dm.unipi.it/astdys/}}. Observations suffer from several sources of systematic errors that can be crucial for asteroid orbit determination \citep{Carpino2003}. Zonal errors of stellar catalogues are one of the systematic errors for observations. \cite{Chesley2010} proposed a method to remove the bias for several stellar catalogues. Bias corrections are given for each observation in the AstDyS database. Finally, the outliers are eliminated thanks to the procedure described in \cite{Carpino2003}. 

\subsection{Fitting and weighting processes}\label{Ss:fit}
The fitting process consists in the determination of seven parameters (the six components of the position and velocity at a given date and the drift in semi-major axis $\dot{a}$ or the $A_2$ parameter defined in Sect.~\ref{Ss:yarko}) that minimises the residuals (observed positions minus computed positions).
This determination uses a Levenberg-Marquardt algorithm by iterative corrections of each parameter. For each iteration, the corrections to apply are determined thanks to the equations of variation and the least square method \citep[LSM; for more information, see for example][]{Desmars2009b}. In the LSM, a weighting matrix $V_{obs}$ is required and usually considered as a diagonal matrix where the diagonal components are $\epsilon_i^2=1/\sigma_i^2$ where $\sigma_i^2$ is the estimated variance of the observation $i$. \\ 

In orbit determination, the main difficulty is to give an appropriate weight to each observation. \cite{Carpino2003} and \cite{Chesley2010} discussed  weighting schemes and showed the orbit determination improvement by weighting observations, according to the observatory and the stellar catalogue used for the reduction. 

The AstDyS database provides the weight for each observation. As a general rule, we use the value given in AstDyS, except for specific cases. As in \cite{Farnocchia2013}, we consider that observations before 1950 have a weight of 10 arsec, and observations from 1950 to 1990 a weight of 3~arcsec. These observations are mainly photographic plates affected by low sensitivity and reduced with less accurate stellar catalogue. The other case is when observations have been carefully reduced again, as in \cite{Vok2008}, with four precovery observations of (152563) 1992BF in January 1953.

\section{Detection of the drift}\label{S:results}
In this section, we try to detect a drift in the semi-major axis of NEAs. We first study the consistency of the drift and its uncertainty determined by LSM by comparing with statistical methods for three representative NEAs (Sect.~\ref{Ss:accuracy}). We also look after the robustness of the drift we determined (Sect.~\ref{Ss:stability}) and we present the drift in semi-major axis for NEAs (Sect.~\ref{Ss:numNEAs}~\&~\ref{Ss:selectedNEAs}) and a comparison with the previous studies (Sect.~\ref{Ss:comp}). Finally, we discuss the constrains in some physical parameters of the asteroids given by the drift (Sect.~\ref{Ss:discussion}).

\subsection{Estimation of uncertainty}\label{Ss:accuracy}
The fitting process gives the orbital elements (more specifically the state vector) of the asteroid and the drift in the semi-major axis. The LSM also provides the covariance matrix $\Lambda$ and the standard deviation of each parameter is given by the root square of diagonal elements. Thus, for the drift in semi-major axis, the standard deviation is given by $\sqrt{\Lambda_{77}}$. This can be considered as a measure of the uncertainty of the drift value $\sigma_{\dot{a}}$. According to LSM theory, the uncertainty of the drift only depends on the date of observations and the weights given to these observations. In particular, it does not depend on the observation coordinates whereas the drift value depends on the date, the weights, and the coordinates of the observations. This aspect is  important because we can study the evolution of the drift uncertainty related to the precision (or the weight) of the observations without the knowledge of the coordinates themselves and thus we can determine the drift uncertainty with simulated observations (see Sect.~\ref{S:datamining}~\&~\ref{S:gaia}).  

To verify if this measure is a good estimation of the uncertainty, we compare $\sigma_{\dot{a}}$ with accuracies obtained by two statistical methods. In this context, we apply a Monte Carlo process on the observations and a bootstrap re-sampling as described in \cite{Desmars2009b}. The Monte Carlo method on the observations (MCO) comes from a technique developed by \cite{Virtanen2001} and consists of adding a random noise on each observation and then fitting the dynamical model to the new set of observations, giving a new drift value. The bootstrap re-sampling (BR) was developed by \cite{Efron1979} and consists in creating a new set of observations by sampling the original set of observations with a draw with replacement. The new set of observations has the same number of elements as the original set. Consequently, some observations can appear several times whereas others do not appear at all. The dynamical model is then fitted to the new set, giving a new drift value. 

For both methods, the process is repeated many times and a measure of the uncertainty is given by the standard deviation of all the drift values. We consider three significative NEAs:
\begin{enumerate}
\item (2340) Hathor for which we determined $\dot{a}=(-14.312\pm3.297)\times 10^{-4}$~au/My, represents NEAs whose signal to noise ratio (S/N) is greater than 3;
\item (138852) 2000WN10 with $\dot{a}=(15.025\pm7.159)\times 10^{-4}$~au/My, represents NEAs with an average S/N;
\item (153792) 2001VH75 with $\dot{a}=(12.812\pm165.373)\times 10^{-4}$~au/My, represents NEAs with very bad detection of the drift.
\end{enumerate}

For both methods (MCO and BR), we generate 1000 samples. Table~\ref{T:MCOBR} presents the value of the drift and the standard deviation obtained with the fitting process (LSM) and  the two different methods (MCO and BR).

\begin{table*}[htdp]
\begin{center}
\caption{Drift in semi-major axis and its standard deviation (in $10^{-4}$~au/My) obtained with three methods for (2340) Hathor, (138852) 2000WN10, and (153792) 2001VH75.}\label{T:MCOBR}
\begin{tabular}{lccc}
\hline
\hline
Asteroid & (2340) Hathor  & (138852) 2000WN10 & (153792) 2001VH75 \\
\hline
LSM  &      $-14.312\pm3.297$  & $15.025\pm7.159$  & $12.812\pm165.373$\\
MCO  &      $-14.417\pm3.294$  & $16.612\pm6.622$  & $18.183\pm158.107$\\
BR   &      $-14.497\pm1.798$  & $14.998\pm3.747$  & $25.342\pm185.706$\\
\hline
\end{tabular}
\end{center}
\end{table*}

The estimations of the uncertainty of the drift are equivalent for LSM and MCO. With the bootstrap method, we have slight differences in particular for NEAs with a bad detection of the drift. Actually, LSM and MCO make use of equivalent assumptions (independence of observations, Gaussian errors of observations) whereas the BR makes use of minimal assumptions (independence of observations only). For NEAs with a good S/N, which is one of our criterions of selection (see Sect.~\ref{Ss:selectedNEAs},) the three methods give similar results. The standard deviation $\sigma_{\dot{a}}$ determined with LSM is a good estimator of the drift in semi-major axis uncertainty and can be used as a measure of the uncertainty.

\subsection{Solution stability}\label{Ss:stability}
The stability of the solution can be tested with a random holdout method. It consists of randomly extracting a percentage of the observations,  fitting the model with these observations, and repeating the process several times. The changes in the fitted parameter gives information on the stability. In this section, we use the same representative NEAs as in Sect.~\ref{Ss:accuracy} and we randomly remove 10\% of the initial set of observations. The process is repeated 1000 times. Table~\ref{T:RHO} presents the mean ($\mu$) and the standard deviation ($\sigma$) of the drift in semi-major axis $\dot{a}$ and its uncertainty $\sigma_{\dot{a}}$, obtained with random holdout method for (2340) Hathor, (138852) 2000WN10, and (153792) 2001VH75.

\begin{table}[htdp]
\begin{center}
\caption{Statistics of drift in semi-major axis $\dot{a}$ and its standard deviation $\sigma_{\dot{a}}$ (in $10^{-4}$~au/My)  obtained with random holdout method for (2340) Hathor, (138852) 2000WN10, and (153792) 2001VH75.}\label{T:RHO}
\begin{tabular}{lcccc}
\hline
\hline
         & \multicolumn{2}{c}{$\dot{a}$}  & \multicolumn{2}{c}{$\sigma_{\dot{a}}$} \\
NEA & $\mu$  & $\sigma $ & $\mu$  & $\sigma$ \\
\hline
(2340) Hathor     & -14.358  &   0.654  &   3.495  &   0.129 \\
(138852) 2000WN10  & 15.089  &   1.254  &   7.473  &   0.084 \\     
(153792) 2001VH75  & 14.223  &  37.476  & 176.367  &   6.120 \\   
\hline
\end{tabular}
\end{center}
\end{table}

The means for $\dot{a}$ and $\sigma_{\dot{a}}$ have to be compared with values given in Table~\ref{T:MCOBR} with LSM. 
By removing 10\% of the initial set, the values obtained for the drift and its uncertainty do not vary too much. It shows that the drift in semi-major axis $\dot{a}$ we detected and its uncertainty $\sigma_{\dot{a}}$ are robust and stable, even for inaccurate determination of $\dot{a}$.

\subsection{Drift determination for numbered NEAs}\label{Ss:numNEAs}
\begin{figure}[h!] 
\centering 
\includegraphics[width=\columnwidth]{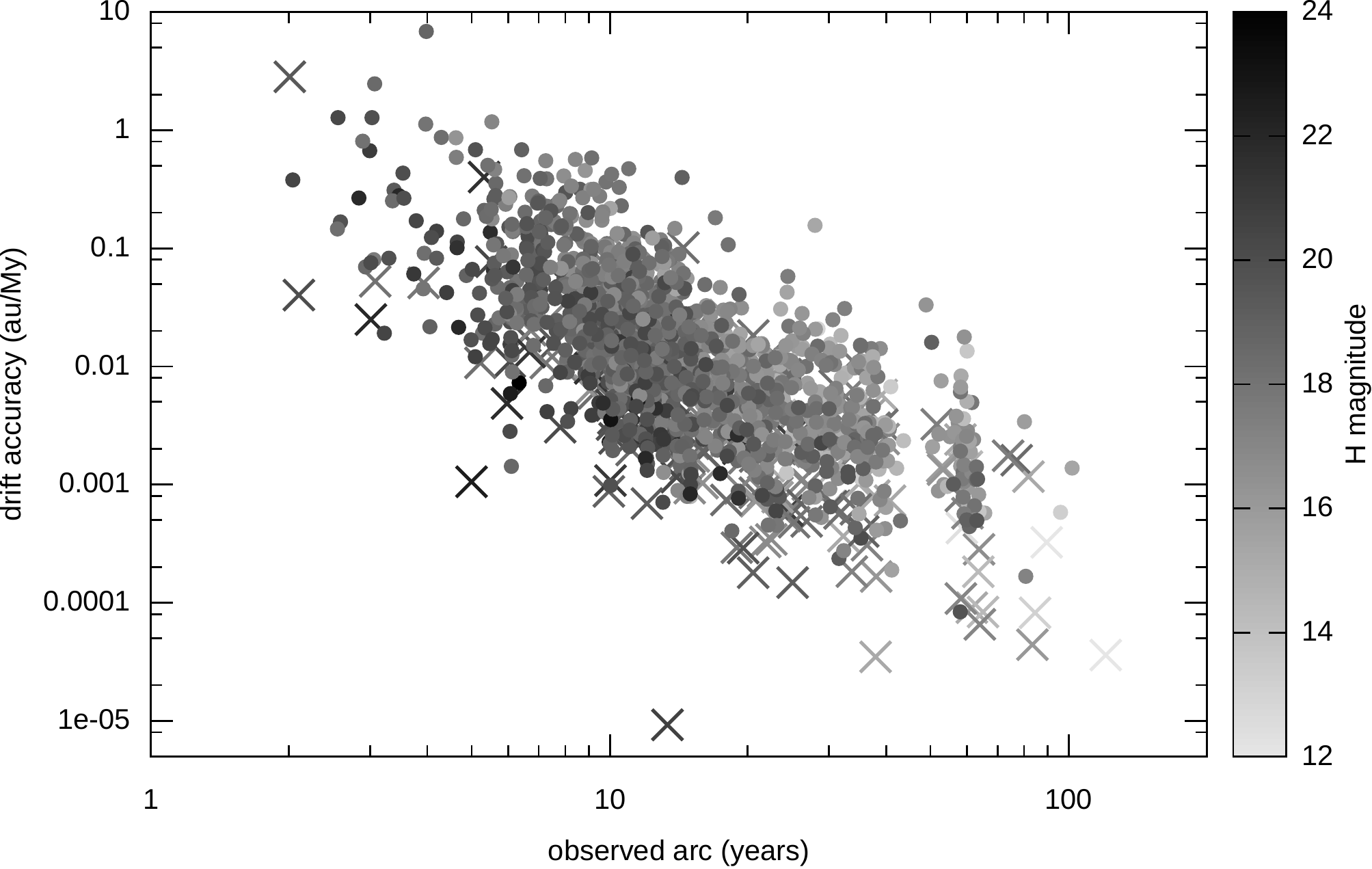}
\caption{Observed arc, uncertainty of the drift in semi-major axis (in au/My), and absolute magnitude $H$ for the 1629 numbered NEAs. NEAs with radar measurements are represented with crosses and NEAs without radar measurements with bullets.}\label{F:arc_drift}
\end{figure}

In this section, we try to determine a drift in semi-major axis for the 1629 numbered NEAs as they have been observed over several oppositions.
Figure~\ref{F:arc_drift} represents the length of the observed arc, the uncertainty of the drift in semi-major axis, and the number of observations for the 1629 numbered NEAs. 

For most NEAs, the uncertainty of the drift we detected is greater than $10^{-3}$~au/My, leading to a bad S/N and inconsistent detections. As the order of magnitude for the drift is about $10^{-3}-10^{-4}$~au/My \citep{Vok2000}, an uncertainty of the same order is at least required for a consistent determination of the drift. 

Figure~\ref{F:arc_drift} shows that a long period of observations (at least 20 years) is necessary to have a small uncertainty of the drift. Nevertheless, with radar measurements, the period of observations can be smaller because they lead to important constraints in the orbit determination \citep[see for example][]{Desmars2013} and in particular for the detection of Yarkovsky acceleration \citep[see the case of (6489) Golevka in][]{Chesley2003}. As the length of orbital arc is longer for bright NEAs, we  do not notice a clear correlation between $H$ magnitude and drift uncertainty.

\subsection{Drift determination for selected NEAs}\label{Ss:selectedNEAs}
We try to detect a drift in semi-major axis for the 10953 NEAs (1629 numbered and 9324 unnumbered) as of 12 May 2014. To select reliable detections of drift in semi-major axis, we adopt two filters.

The first filter is the signal to noise ratio defined as S/N$=|\dot{a}|/\sigma_{\dot{a}}$. \cite{Farnocchia2013} used S/N$\ge3$ as a threshold to define reliable detections but they also considered lower values for some specific cases. This threshold appears restrictive because a detection with S/N$\sim2$ also leads to consistent drifts. In contrast, \cite{Nugent2012} used S/N$\ge1$ ,which seems to be a nonrestrictive threshold. If we assume a Gaussian distribution for the drift value and if S/N=1, the probability of detecting a drift value with an opposite sign from the real value (for example a positive drift where the real drift is negative or, conversely, a negative drift where the real drift is positive) is about 15.9\%, whereas if S/N=2, this probability is 2.3\%. In that context, we adopt S/N$\ge2$ as an appropriate filter. 

The second filter is to avoid spurious detections: for example, the detection of a strong drift for large NEAs. We estimated a maximum value for the drift in the semi-major axis according to the absolute magnitude $H$ of the NEA,

\begin{eqnarray}
\dot{a}_{max} = \left|\left(\frac{da}{dt}\right)_d\right|_{max} & = & 10^{0.2H-6.85} \text{ au/My}\label{eqmax}
.\end{eqnarray}

The maximum value assumes average values of semi-major axis, albedo, and bulk density for NEA according to \cite{Chesley2002}. The full details of the estimation are given in Appendix~\ref{SAppendix}. This parameter remains an approximation and can be underestimated by a factor of 3 for NEA with low density and by a factor of 5 for NEAs with high eccentricity \citep{Vok2000}.
The ratio $\xi=|\dot{a}|/\dot{a}_{max}$ can help to filter spurious detection. The parameter $\xi$ is expected to be smaller than 1 but by considering the approximate value of the maximum drift, we select NEAs for which $\xi \le 3$.

Finally, Table~\ref{T:value1} presents the selected NEAs, for which S/N$\ge 2$ and $\xi \le 3$, with the drift in semi-major axis $\dot{a}$, its uncertainty $\sigma_{\dot{a}}$, the $A_2$ parameter and its uncertainty $\sigma_{A_2}$, the signal to noise ratio S/N, the absolute magnitude $H$, the diameter $D$ (from EARN database\footnote{\url{http://earn.dlr.de/nea/}}), the semi-major axis (in au), the eccentricity, the $\xi$ ratio, and the time span of observations. An additional asterisk indicates that radar measurements are available for this NEA. We also add three NEAs with $\sigma_{\dot{a}}\le10^{-4}$~au/My because as discussed later, this sort of a precision indicates a drift close to zero, which provides an initial constraint. 

Most of $\xi$ values are close to 1, indicating that a drift detection is currently possible when the drift rate is close to its maximum value. Moreover, $\xi$ close to 2 or 3 are associated with a high eccentricity.

\begin{table*}[htdp]
\caption{Drift in semi-major axis and its uncertainty, $A_2$ parameter and its uncertainty, S/N, $H$ magnitude, diameter, semi-major axis, eccentricity, $\xi$ ratio, and orbital arc for 43 selected NEAs with S/N$\ge2$ and $\xi \le 3$ and 3 NEAs with $\sigma_{\dot{a}}\le10^{-4}$~au/My (see text).} \label{T:value1}
\begin{center}
\begin{tabular}{rlrrrrrrrrrrl}
\hline
\hline
\multicolumn{2}{c}{NEA}  & \multicolumn{1}{c}{$\dot{a}$} & \multicolumn{1}{c}{$\sigma_{\dot{a}}$} & \multicolumn{1}{c}{$A_2$} & \multicolumn{1}{c}{$\sigma_{A_2}$} & \multicolumn{1}{c}{S/N} & \multicolumn{1}{c}{$H$} & \multicolumn{1}{c}{$D$} & \multicolumn{1}{c}{$a$} & \multicolumn{1}{c}{$e$} & \multicolumn{1}{c}{$\xi$} & \multicolumn{1}{c}{time-span}\\
 & & \multicolumn{2}{c}{($10^{-4}$au/My)} & \multicolumn{2}{c}{($10^{-15}$au/d$^2$)} & & & \multicolumn{1}{c}{(m)} & \multicolumn{1}{c}{(au)} & & & \\
\hline
\hline
101955          & Bennu      &    -18.611 &      0.093 &    -44.581 &      0.222 & 200.6 &  20.2&     484 &     1.13 &     0.20 &     0.99 & 1999-2013* \\
152563          & 1992BF     &    -11.297 &      0.839 &    -23.482 &      1.744 &  13.5 &  19.8&     510 &     0.91 &     0.27 &     0.94 & 1953-2011  \\
                & 2005ES70   &    -69.056 &      7.823 &   -120.810 &     13.687 &   8.8 &  23.8&      -- &     0.76 &     0.39 &     0.89 & 2005-2013  \\
3908            & Nyx        &     11.545 &      1.754 &     29.781 &      4.525 &   6.6 &  17.3&    1000 &     1.93 &     0.46 &     2.86 & 1980-2014* \\
4179            & Toutatis   &     -2.087 &      0.363 &     -4.718 &      0.821 &   5.7 &  15.3&    2800 &     2.53 &     0.63 &     1.37 & 1976-2014* \\
                & 2006CT     &    -49.077 &      8.607 &   -114.620 &     20.101 &   5.7 &  22.2&      -- &     1.10 &     0.23 &     1.22 & 1991-2014* \\
152664          & 1998FW4    &     15.537 &      2.910 &     27.818 &      5.210 &   5.3 &  19.7&      -- &     2.52 &     0.72 &     1.28 & 1994-2013* \\
2062            & Aten       &     -5.572 &      1.084 &    -12.469 &      2.426 &   5.1 &  17.1&    1300 &     0.97 &     0.18 &     1.56 & 1955-2014* \\
1862            & Apollo     &     -2.314 &      0.491 &     -4.537 &      0.963 &   4.7 &  16.4&    1400 &     1.47 &     0.56 &     1.02 & 1930-2014* \\
                & 2009BD     &   -418.170 &     88.640 &   -986.640 &    209.140 &   4.7 &  28.1&       4 &     1.01 &     0.04 &     0.67 & 2009-2011  \\
                & 2007TF68   &    -98.134 &     21.758 &   -254.770 &     56.487 &   4.5 &  22.7&      -- &     1.41 &     0.26 &     2.02 & 2002-2012  \\
29075           & 1950DA     &     -2.729 &      0.656 &     -6.217 &      1.495 &   4.2 &  17.6&    2000 &     1.70 &     0.51 &     0.75 & 1950-2014* \\
                & 2004KH17   &    -42.423 &     10.228 &    -63.288 &     15.258 &   4.1 &  21.9&     197 &     0.71 &     0.50 &     1.26 & 2004-2013* \\
2340            & Hathor     &    -14.131 &      3.529 &    -24.377 &      6.087 &   4.0 &  20.2&     300 &     0.84 &     0.45 &     1.00 & 1976-2012  \\
37655           & Illapa     &    -11.351 &      2.925 &    -14.090 &      3.631 &   3.9 &  18.0&      -- &     1.48 &     0.75 &     2.23 & 1994-2013* \\
54509           & YORP       &    -36.240 &     10.581 &    -80.829 &     23.600 &   3.4 &  22.6&     100 &     1.00 &     0.23 &     0.79 & 2000-2005* \\
2100            & Ra-Shalom  &     -4.502 &      1.318 &     -7.827 &      2.291 &   3.4 &  16.1&    2240 &     0.83 &     0.44 &     1.89 & 1975-2013* \\
4581            & Asclepius  &    -19.683 &      6.031 &    -40.896 &     12.532 &   3.3 &  20.7&      -- &     1.02 &     0.36 &     1.00 & 1989-2013* \\
1620            & Geographos &     -2.708 &      0.913 &     -6.315 &      2.129 &   3.0 &  16.5&5000$^1$ &     1.25 &     0.34 &     1.77 & 1951-2013* \\
350462          & 1998KG3    &    -24.853 &      8.343 &    -62.171 &     20.871 &   3.0 &  22.1&      -- &     1.16 &     0.12 &     0.68 & 1998-2013  \\
6489            & Golevka    &     -5.468 &      1.796 &    -13.120 &      4.310 &   3.0 &  19.1& 350$^2$ &     2.52 &     0.60 &     0.62 & 1991-2011* \\
283457          & 2001MQ3    &    -14.692 &      5.027 &    -40.926 &     14.005 &   2.9 &  19.1&     500 &     2.23 &     0.46 &     1.73 & 1951-2011  \\
256004          & 2006UP     &    -95.050 &     35.471 &   -256.300 &     95.646 &   2.7 &  23.0&      -- &     1.59 &     0.30 &     1.64 & 2002-2012  \\
363599          & 2004FG11   &    -28.744 &     10.805 &    -40.589 &     15.257 &   2.7 &  21.0&     152 &     1.59 &     0.72 &     1.28 & 2004-2014* \\
85770           & 1998UP1    &    -16.118 &      5.982 &    -33.419 &     12.403 &   2.7 &  20.5&      -- &     1.00 &     0.34 &     0.95 & 1990-2013  \\
                & 1999SK10   &    -20.989 &      7.650 &    -52.845 &     19.260 &   2.7 &  19.7&     400 &     1.76 &     0.44 &     1.73 & 1999-2014  \\
                & 1995FJ     &     42.330 &     16.182 &     96.248 &     36.795 &   2.6 &  20.5&     300 &     1.08 &     0.27 &     2.69 & 1995-2013  \\
                & 2001QC34   &    -32.160 &     12.255 &    -77.598 &     29.570 &   2.6 &  20.1&      -- &     1.13 &     0.19 &     2.22 & 2001-2014  \\
99907           & 1989VA     &     11.931 &      4.520 &     15.498 &      5.871 &   2.6 &  18.0&     550 &     0.73 &     0.59 &     2.33 & 1989-2012  \\
164207          & 2004GU9    &    -67.288 &     26.497 &   -155.570 &     61.261 &   2.5 &  21.1&     163 &     1.00 &     0.14 &     2.81 & 2001-2014  \\
                & 1999MN     &     28.367 &     11.432 &     30.576 &     12.322 &   2.5 &  20.6&      -- &     0.67 &     0.67 &     1.13 & 1999-2010  \\
230111          & 2001BE10   &    -17.126 &      6.911 &    -31.614 &     12.758 &   2.5 &  19.1&     400 &     0.82 &     0.37 &     1.79 & 2001-2013* \\
                & 2005EY169  &    -39.349 &     17.142 &   -100.220 &     43.662 &   2.3 &  22.1&      -- &     1.31 &     0.23 &     1.06 & 2005-2014  \\
388189          & 2006DS14   &    -30.247 &     13.204 &    -58.690 &     25.620 &   2.3 &  20.5&     315 &     0.86 &     0.34 &     1.79 & 2002-2014  \\
                & 2004BG41   &   -245.090 &    110.220 &   -573.400 &    257.850 &   2.2 &  24.4&      -- &     2.52 &     0.61 &     2.25 & 2004-2011  \\
390522          & 1996GD1    &    -37.800 &     17.403 &    -84.884 &     39.080 &   2.2 &  20.6&      -- &     1.19 &     0.35 &     2.41 & 1996-2014  \\
3361            & Orpheus    &      5.220 &      2.372 &     12.109 &      5.502 &   2.2 &  19.0&     348 &     1.21 &     0.32 &     0.54 & 1982-2013  \\
154590          & 2003MA3    &    -46.441 &     21.733 &    -96.389 &     45.108 &   2.1 &  21.7&      86 &     1.11 &     0.40 &     1.53 & 1998-2012  \\
138852          & 2000WN10   &     15.025 &      7.079 &     32.239 &     15.190 &   2.1 &  20.2&      -- &     1.00 &     0.30 &     1.02 & 2000-2013  \\
162361          & 2000AF6    &     18.534 &      9.037 &     33.979 &     16.569 &   2.1 &  20.1&      -- &     0.88 &     0.41 &     1.26 & 1991-2014  \\
377097          & 2002WQ4    &    -10.360 &      4.971 &    -23.635 &     11.342 &   2.1 &  19.6&      -- &     1.96 &     0.56 &     0.92 & 1950-2014  \\
350523          & 2000EA14   &     48.659 &     24.108 &    116.120 &     57.534 &   2.0 &  21.1&      -- &     1.12 &     0.20 &     2.08 & 2000-2013  \\
7336            & Saunders   &     19.567 &      9.751 &     53.736 &     26.780 &   2.0 &  19.0&     600 &     2.30 &     0.48 &     2.46 & 1982-2010  \\
\hline
1685            & Toro       &     -1.225 &      0.826 &     -2.732 &      1.842 &   1.5 &  14.0&    3750 &     1.37 &     0.44 &     1.19 & 1948-2013* \\
1627            & Ivar       &      0.256 &      0.820 &      0.694 &      2.221 &   0.3 &  12.6&    8370 &     1.86 &     0.40 &     0.44 & 1929-2014*  \\  
433             & Eros       &     -0.283 &      0.361 &     -0.765 &      0.977 &   0.8 &  10.3&   23300 &     1.46 &     0.22 &     1.40 & 1893-2014* \\
\hline
\hline
 & \multicolumn{12}{l}{$^1$ (1620) Geographos size is 5x2x1.5km.}\\
 & \multicolumn{12}{l}{$^2$ (6489) Golevka size is 350x250x250m.}
\end{tabular}
\end{center}
\end{table*}

\subsection{Comparison with previous results}\label{Ss:comp}
\begin{figure}[h!] 
\centering 
\includegraphics[width=\columnwidth]{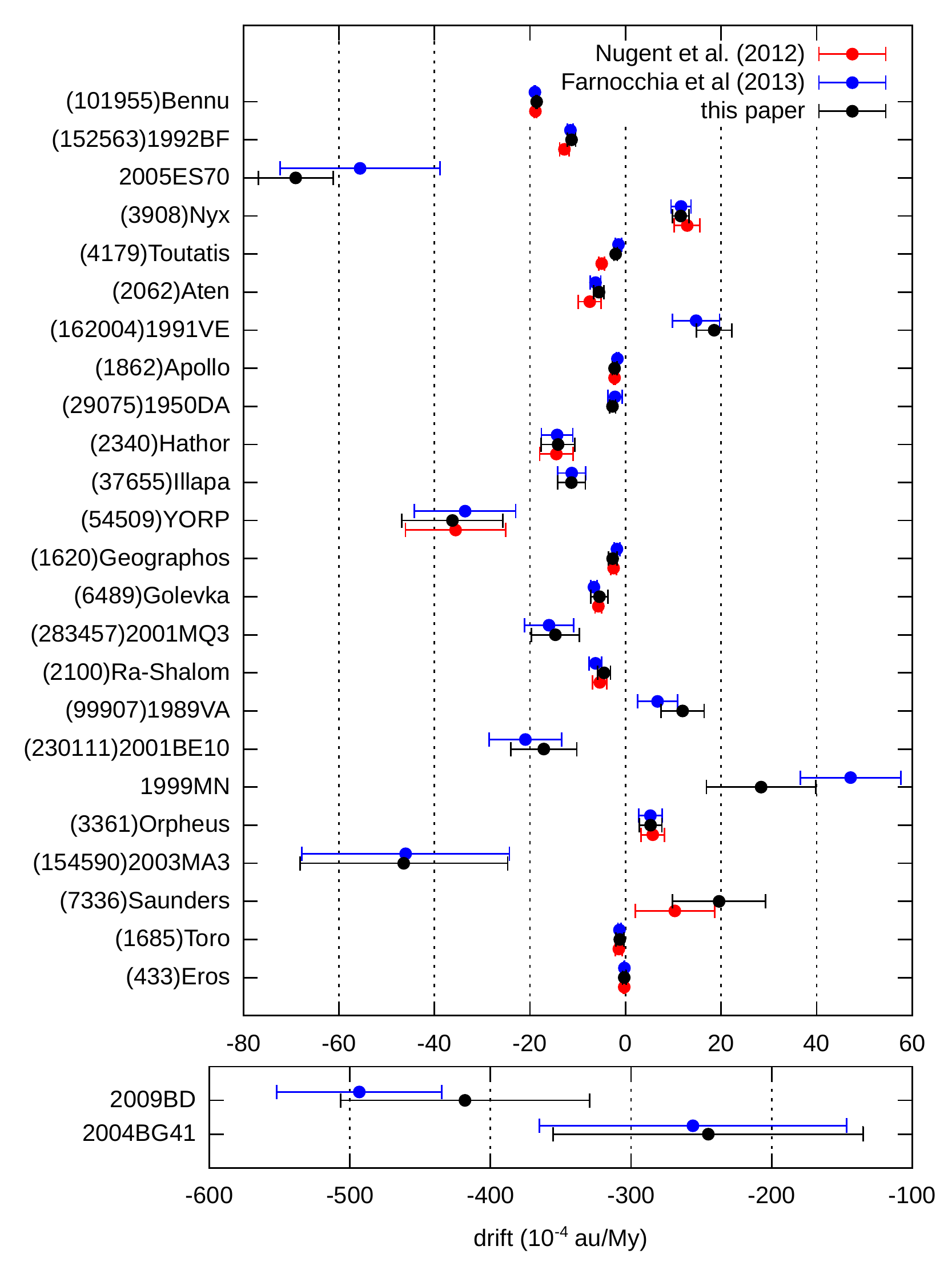}
\caption{Comparison of the drift in semi-major axis we obtained and that of previous studies \citep{Farnocchia2013,Nugent2012}.}\label{F:comparison}
\end{figure}

As already mentioned, previous studies \citep{Farnocchia2013,Nugent2012} also detected a drift in semi-major axis for several NEAs. We compare our values with those of  previous studies (Fig.~\ref{F:comparison}). 

Our results are in good agreement with the results previously published. The difference can be mainly explained by additional observations. For example, \cite{Nugent2012} used observations until 2012, \cite{Farnocchia2013} until 2013, and we used observations until May 2014. 
A different threshold for rejection may also explain some differences. For example, the drift for 1999MN is sensitive to the rejection threshold because by using a less restrictive threshold, we have $\dot{a}=(43.2\pm10.7)\times 10^{-4}$~au/My , which is closer to the drift obtained by \cite{Farnocchia2013}. This may indicate a problem with astrometry for this object that should be remeasured. 

Finally, considering the 1-$\sigma$ uncertainty, the greater discrepancy with previous results is for (101955)Bennu, with a difference of about 4~$\sigma$. An attentive control of the astrometry and radar measurements do not allow us to explain this discrepancy. 

\subsection{Discussion}\label{Ss:discussion}
The drift in semi-major axis can be related to physical parameters. For example, as the diurnal component is dominant, \cite{Vok2008} provides a relation between the drift, the obliquity of spin axis $\gamma$, the bulk density $\rho_b$, the diameter $D$ , and the diurnal thermal parameter $\Theta$, i.e.
\begin{equation}\label{E:phys}
\frac{da}{dt}\propto\frac{\cos \gamma}{\rho_b D}\frac{\Theta}{1+\Theta+0.5\Theta^2}
.\end{equation}

On the one hand, the knowledge of physical parameters can provide expected drift rates. For example, \cite{Nugent2012b} used diameters and albedo from WISE data to predict drift rates due to the Yarkovsky effect for 540 NEAs.  
On the other hand, the knowledge of $da/dt$ can constrain physical parameters $D, \rho \text{ , or } \Theta$. In particular, if one of these parameters is already known, it would be helpful for the determination of other unknown parameters. 

We  also note the relation between the diameter of the NEA and its drift value. Diameter, absolute magnitude, and albedo of NEAs are available on the EARN database\footnote{\url{http://earn.dlr.de/nea/}} and when they are not available, we consider the diameter (in km) given by the relation with the absolute magnitude $H$ and the albedo $p_v$ and a parameter $K=1329$~km \citep{Pravec2007}, i.e. 

\begin{equation}\label{E:diameter}
D\sqrt{p_v}=K \times 10^{-H/5}
.\end{equation}

\begin{figure}[h!] 
\centering 
\includegraphics[width=\columnwidth]{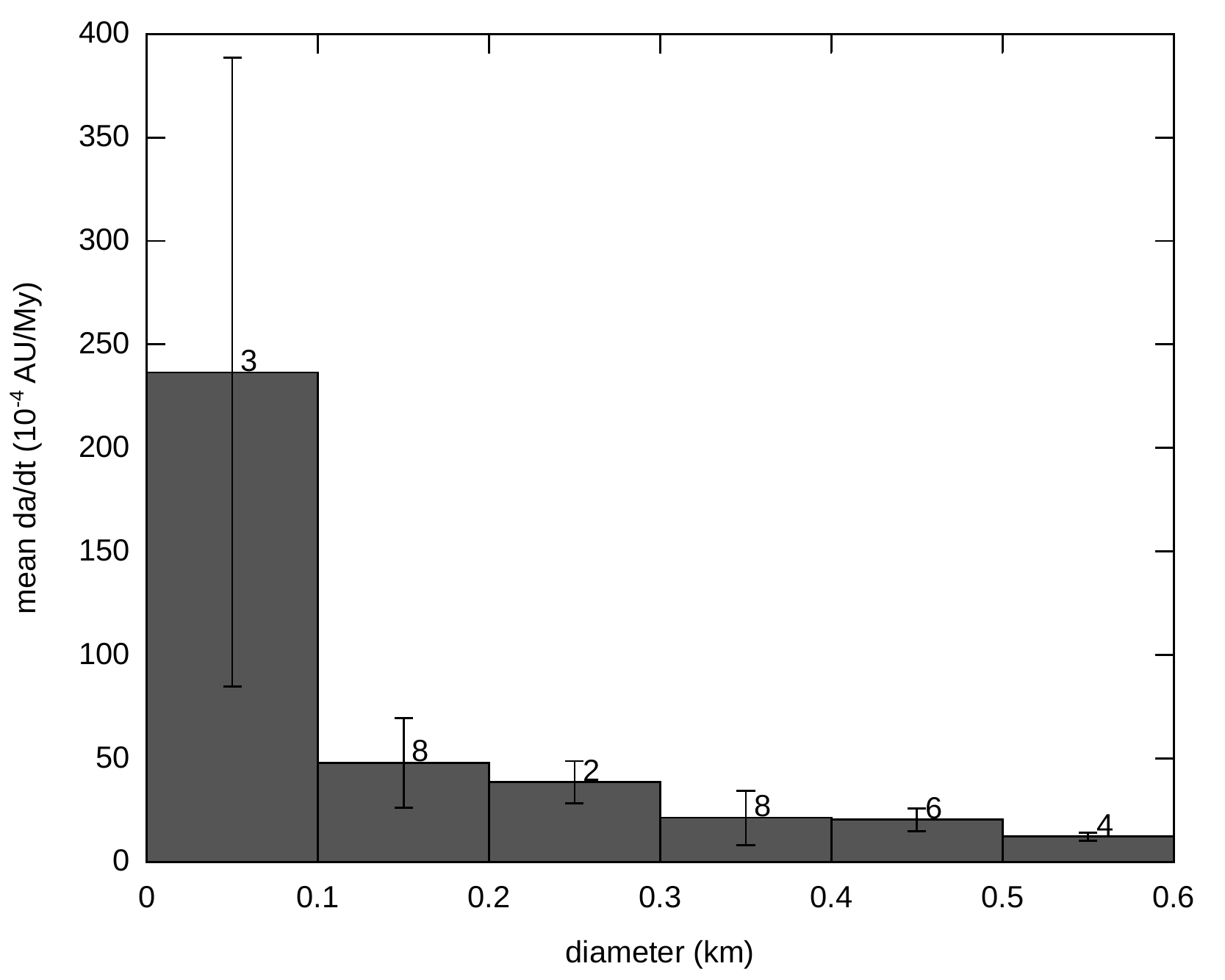}
\caption{Distribution of diameter and mean value of $\dot{a}$ for NEAs smaller than 0.6~km. Number of component is indicated above each bar of the histogram and the standard deviation of the mean is represented by error bar.}\label{F:diamdadt}
\end{figure}

As most of the NEAs we selected have a diameter smaller than 0.6~km, we can obtain statistics for the distribution of the diameter and the mean value of the drift (in absolute value). For NEAs bigger than 0.6~km, we do not have enough objects to obtain consistent statistics. Figure~\ref{F:diamdadt} represents this distribution for small NEAs ($D\le0.6$~km). The number of components of each bar of the histogram is also indicated. Even if the drift in semi-major axis does not depend only on the diameter (see Eq.~\ref{E:phys}), we note the expected $1/D$ dependence of $|\dot{a}|$.

Another interesting result can be found in the mechanism of NEA transportation described in \cite{Bottke2002} because one of the consequences of this mechanism is an excess of retrograde rotators in the NEA population. According to \cite{LaSpina2004}, 67\% of the NEAs should be retrograde. 

Actually, in equation~\ref{E:phys}, the sign of $\dot{a}$ allows us to distinguish retrograde NEAs (negative drift) and prograde NEAs (positive drift). For the 43 selected NEAs from Table~\ref{T:value1}, we found 77\% of retrogade rotators, which is close to the distribution given in \cite{LaSpina2004}.

\section{The precovery context}\label{S:datamining}
As a long period of observations is necessary to detect an accurate drift in semi-major axis, data mining and precovery observations can appear very important in that context. The case of (152563) 1992BF studied in \cite{Vok2008} is a significative example because four precoveries were available and reduced with a modern catalogue  to refine astrometry and then orbit determination. 

Nowadays, photographic plates used until the 1990s can be digitised and reduced with modern techniques and modern stellar catalogues \citep{Robert2011} as the current observations on CCD frames. In this context, old observations can have a precision equivalent to current observations. 

This section aims to investigate the case of NEAs with existing precoveries for which a new reduction of these observations would help to better determine the drift in  semi-major axis. Then, by studying two specific NEAs, we highlight favourable periods that allow a better uncertainty of the drift, showing that data mining has to be considered  a great opportunity in the detection of drift.
 
\subsection{Existing precoveries}
In this section, we identify NEAs for which precovery observations have been isolated but not used for orbit determination because of their inaccuracy. These observations allow us to extend the orbital arc for several tens of years and, in the case of a new reduction, they would allow for  a better determination of the drift in semi-major axis, or at least a smaller uncertainty. Depending on the quality of the precovery observations, we consider two different precisions for the precoveries with a new reduction: 300~mas\footnote{mas is for milli-arcsec.} corresponding to accurate observations, and 1~arcsec corresponding to low-quality observations. We determine $\sigma_{\dot{a}}$ considering both cases. In any case, the new reduction will decrease the uncertainty in the drift, but to select NEAs for which the new reduction would be helpful for the drift detection, we filter NEAs with the ratio $\dot{a}_{max}/\sigma_{\dot{a}}$, where $\sigma_{\dot{a}}$ is obtained considering a 300~mas precision for precovery. In fact, in the case in which the drift is close to its maximum $\dot{a}_{max}$, this ratio gives an approximate S/N obtained with the new reduction.   
Table~\ref{T:precovery} presents 14 NEAs with a ratio $\dot{a}_{max}/\sigma_{\dot{a}}\ge1$ and indicates the year of the precovery, the length of additional arc (the time between the first observation used in orbit determination and the precovery), the number of precoveries, the current uncertainty in the drift in semi-major axis and the expected uncertainty of the drift in the case the precoveries can be reduced with an precision of 1~arcsec and 300~mas and the ratio $\dot{a}_{max}/\sigma_{\dot{a}}$.  

\begin{table*}[htdp]
\caption{NEAs with existing precoveries for which a new reduction will be helpful (see text). The year of the first precovery, the additional extended arc, the number of available precoveries, and the current and expected drift uncertainty obtained for various precision of precoveries, are presented.} \label{T:precovery}
\begin{center}
\begin{tabular}{rlcrrrrrr}
\hline
\hline
\multicolumn{2}{c}{Asteroid}  & Year of & Additional arc & Number of & Current $\sigma_{\dot{a}}$ & Expected $\sigma_{\dot{a}}$ & Expected $\sigma_{\dot{a}}$ & $\dot{a}_{max}/\sigma_{\dot{a}}$ \\
 & & 1\textsuperscript{st} precovery &  &  precoveries & & (1~\arcsec) & (300~mas)  &  \\
 & &  & (years) &    & \multicolumn{3}{c}{($10^{-4}$au/My)} &  \\
\hline
\hline
162421      & 2000ET70    & 1977 &  23.050 &  2 &     16.8 &      1.7 &      0.7 &     8.04 \\
4179        & Toutatis    & 1934 &  42.292 &  2 &      0.4 &      0.4 &      0.3 &     4.43 \\
138175      & 2000EE104   & 1999 &   0.923 &  5 &     11.6 &      5.8 &      4.5 &     3.46 \\
7350        & 1993VA      & 1963 &  22.563 &  3 &      7.7 &      2.0 &      1.5 &     2.37 \\
66400       & 1999LT7     & 1987 &  12.134 &  2 &     19.1 &      7.0 &      4.5 &     2.29 \\
5797        & Bivoj       & 1953 &  26.253 &  1 &      4.3 &      3.6 &      3.4 &     2.28 \\
350751      & 2002AW      & 1991 &  10.776 &  2 &     72.9 &     24.2 &      9.8 &     2.01 \\
            & 1999FA      & 1978 &  19.855 &  1 &    127.7 &     10.3 &      9.7 &     1.96 \\
87024       & 2000JS66    & 1979 &  21.328 &  2 &     32.4 &     11.0 &      4.8 &     1.54 \\
            & 2004SV55    & 1983 &  20.997 &  5 &    576.5 &      4.7 &      3.9 &     1.31 \\
7341        & 1991VK      & 1981 &   9.886 &  1 &      3.3 &      3.2 &      2.4 &     1.27 \\
1943        & Anteros     & 1968 &   4.767 &  2 &      1.9 &      1.9 &      1.7 &     1.05 \\
5604        & 1992FE      & 1976 &   8.983 &  2 &     10.0 &      5.5 &      3.5 &     1.01 \\
4450        & Pan         & 1937 &  50.631 &  2 &     10.8 &      4.7 &      3.7 &     1.01 \\
\hline
\hline
\end{tabular}
\end{center}
\end{table*}
Currently, these NEAs have a bad determination of the drift, but a new reduction of the precoveries would help to decrease the uncertainty. Even if the precision of precoveries with the new reduction is 1~arcsec (corresponding to a low-quality observation), we can expect a large improvement of the drift uncertainty especially as the length of orbital arc is increasing. Obviously, the uncertainty of the drift is much smaller as we consider 300~mas for the precision of precoveries.

\subsection{Importance of data mining}
Photographic plates used before the 1990s were used for various purposes and surveys. The field of view covered is usually several degrees and many objects can appear on the plates, in particular, NEAs leading to precovery observations. In this section, we investigate the impact of the date and the precision of the precovery on the drift uncertainty. To measure the importance of the precovery observations, we consider the case of (152563) 1992BF , except with only the 1992--2011 observations (i.e. without the four precoveries of 1953). The idea is to understand why the 1953 precoveries are important and what happen if the precovery date changes.

In this context, we consider one additional precovery observation made between 1950 and 1990 and we determine the uncertainty of the detected drift with this precovery observation. For this, we consider various values of precision for the precovery observation: 3~arcsec, which corresponds to a bad astrometry (inaccurate stellar catalogue, bad quality of photographic plates), 300~mas, which corresponds to mean precision of current observations, and 30~mas , which could correspond to the precision obtained with the future Gaia stellar catalogue (see Sect.~\ref{S:gaia}). 

\begin{figure}[h!] 
\centering 
\includegraphics[width=\columnwidth]{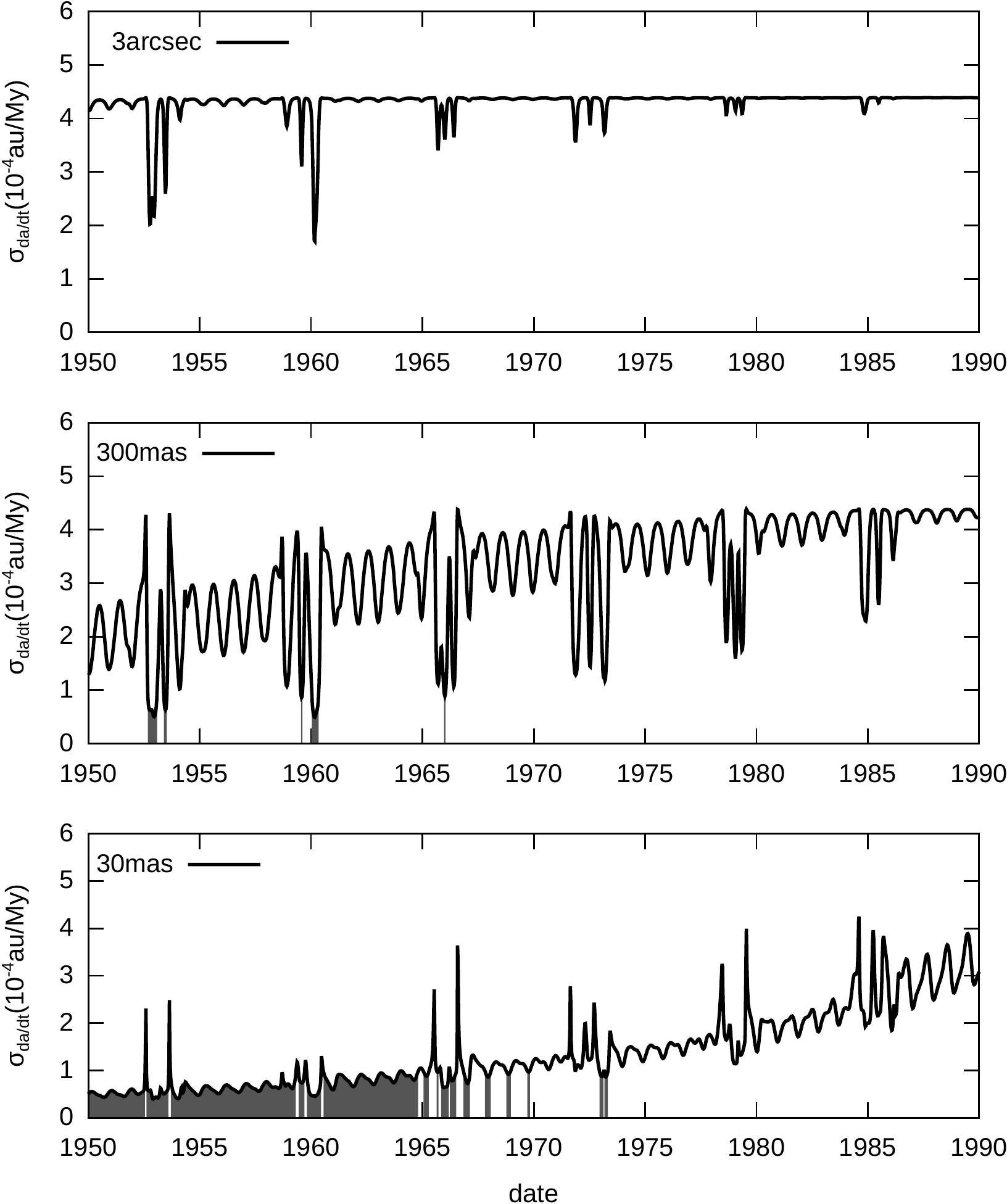}
\caption{Uncertainty of the drift in semi-major axis related to the date of the precovery observations for (152563) 1992BF, with a precision of 3~arcsec, 300~mas , and 30~mas from top to bottom, for the precovery observation. Unit is $10^{-4}$au/My for $\sigma_{\dot{a}}$ in ordinate.}\label{F:vardadt_152563}
\end{figure}

Figure~\ref{F:vardadt_152563} shows the uncertainty of the drift in semi-major axis obtained in that context for (152563) 1992BF. The grey zone under the curve indicates period for which $\sigma_{\dot{a}}$ is smaller than $10^{-4}$au/My, allowing for an accurate detection of the drift. 

If the precovery observation has an precision of 3~arcsec, then most of the time $\sigma_{\dot{a}}$ is close to $4.4\times 10^{-4}$au/My , which is the uncertainty obtained with only 1992--2011 observations. During short period of time (1953, 1960), the uncertainty becomes close to $2\times 10^{-4}$au/My. These periods correspond to close approaches with Earth, when the geocentric distance of the asteroid is smaller than 0.4~au. With a 300~mas precision for the precovery observation, we still have these favourable periods (1953, 1960, 1966) where $\sigma_{\dot{a}}$ becomes smaller than $10^{-4}$au/My, but it  becomes even smaller when the precovery is farther in the past. This is even more the case with a precision of 30~mas for the precovery observation, where $\sigma_{\dot{a}} \le10^{-4}$au/My , whatever the date of precovery observation before 1965 might be. 

Close approaches with Earth appear as good opportunities to improve the uncertainty of the drift in semi-major axis. Indeed, NEAs move faster in the celestial sphere leading to stronger constraints  on their positions and in orbit determinations. Besides, during close approaches, NEAs appear brighter than usual and can leave  traces on photographic plates. 

We repeat the study for (99942) Apophis, which is a NEA with a very close approach to Earth on April 2029, at about 38\,000~km. Currently, we have $\sigma_{\dot{a}}=8.778\times 10^{-4}$~au/My and very small S/N, which is obviously not significant, but provides an initial constraint.

\begin{figure}[h!] 
\centering 
\includegraphics[width=\columnwidth]{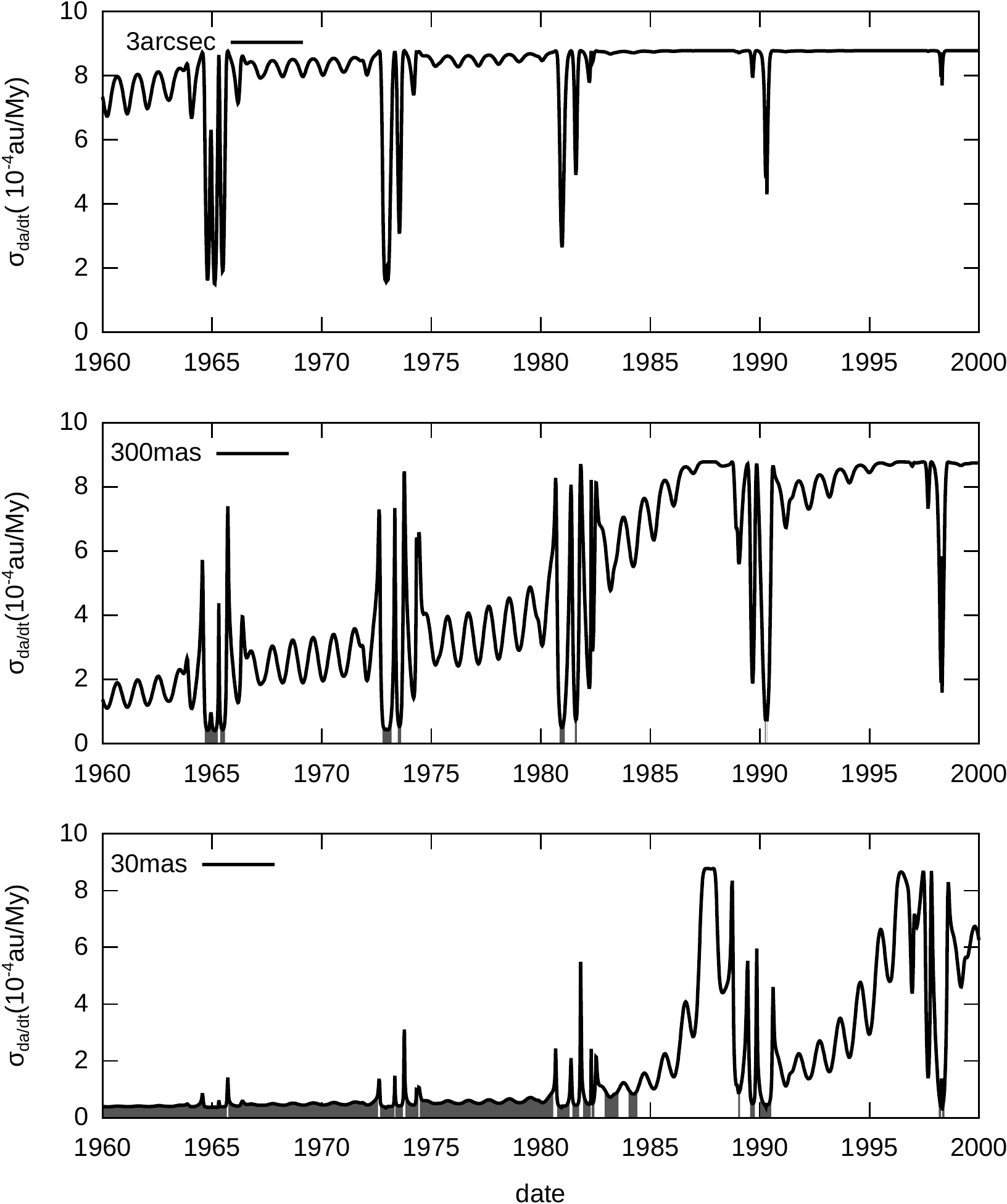}
\caption{Uncertainty of the drift in semi-major axis related to the date of the precovery observations for (99942) Apophis, with a precision of 3~arcsec, 300~mas, and 30~mas from top to bottom, for the precovery observation. Unit is $10^{-4}$au/My for $\sigma_{\dot{a}}$ in ordinate.}\label{F:vardadt_99942}
\end{figure}

Figure~\ref{F:vardadt_99942} shows that the uncertainty of the drift could be reduced by a precovery observation performed during favourable periods (1965, 1972, 1980, 1990), which corresponds to close approach with Earth when a geocentric distance is smaller than 0.1~au. On these occasions, Apophis could have been imaged on photographic plates. Actually, we found a Beijing Observatory plate that covers the position of Apophis on January 1981. Unfortunately, because of its apparent magnitude of about 18, Apophis could not be detected on the plate. Nevertheless, in the case of a positive precovery observation at this date and with a precision of 300~mas, it would be possible to have a drift uncertainty of about $0.8\times 10^{-4}$au/My.

Using astrometric observations, there are two ways to decrease the drift uncertainty: by extending the orbital arc of the NEA (see Fig.~\ref{F:arc_drift}) and during NEA close approaches with Earth. Figures~\ref{F:vardadt_152563}~\&~\ref{F:vardadt_99942} show that the drift uncertainty $\sigma_{\dot{a}}$ is smaller when the geocentric distance is small. During these periods, NEA appears brighter and during  past close approaches, bright NEA could have been imaged on photographic plates. This is the case for several objects, including (152563) 1992BF, for which the photographic plates come from the Palomar Mountain series that was used for the DSS survey. This is an example of an exploited photographic archive, but other observatories in the world may have unexploited archives and could bring new observations to improve orbit determination and drift detection. 

\section{The Gaia mission context} \label{S:gaia}
The space mission Gaia was launched successfully on December 2013. The aim is to have a 3D-map of our Galaxy. In the case of solar system objects, the satellite will enable us to map thousands of main belt asteroids and NEAs down to magnitude 20. Gaia will observe several times each NEA for five years of the mission. \cite{Mouret2011} identified 64 promising NEAs for which Gaia observations could help the detection of a drift in semi-major axis with an uncertainty better than $5\times 10^{-4}$au/My. 

Beyond the Gaia observations themselves, one of the best expected improvements will be the Gaia stellar catalogue allowing for orbital improvement and detection of drift in asteroid semi-major axis. The Gaia catalogue \citep{Mignard2007} will provide unbiased positions of a billion  stars down\ to magnitude 20 and with a precision depending on magnitude (7 $\mu$as at $\le$10 mag; 12-25 $\mu$as at 15 mag and 100-300 $\mu$as at 20 mag). 

With such a precision, new processes of reduction will be necessary, such as propagation models to third order for star proper motions, differential aberration, atmospheric absorption, color of stars, etc. Finally, by including these corrections and considering the limitation due to the seeing, an uncertainty of 10-20 mas can be expected on the position of any asteroid reduced with the Gaia stellar catalogue. 

Moreover, the Gaia stellar catalogue will enable us to reduce old photographic plates. The reduced position will be affected by other errors or limitations, including the distortion or instrument deformations; the granularity of the photographic plate; the digitisation; the exposure time; and the proper motions of stars. Depending on the magnitude and the epoch, the error would be, for example, about 12 mas for 20-mag star and 1.5mas for a 16-mag star at epoch 1950 \citep{Mignard2013}. 

Another problem could be raised from a lack of meta-data in time observations. The time of the observations could be given with a low precision (to the minute, for example), or an error clock in the time of observation could possibly lead to inaccurate or even worse unusable observations.  

In this study, we do not consider clock error, but only errors due to digitisation, photographic plates, or instruments. Assuming a good plate quality, fine grain emulsion, and optimal plate measuring, \cite{Zacharias2013} estimated an error budget of about 10-20 mas for an astrometric position extracted from digitised plate. For these reasons, we assume that the astrometric precision of the new reduction with Gaia stellar catalogue will be 10 mas for observations since 1990 (for mainly CCD observations), 30 mas for 1950--1990 observations, 50 mas for 1920--1950 observations, and 100 mas for observations before 1920 (mainly photographic plates).

To measure the improvement of the drift detection with the Gaia stellar catalogue, we choose several assumptions as to new reduction methods that make sue of the Gaia catalogue. It seems obvious that all of the observations could not be reduced again, thus we assume that only a few would be reduced. As mentioned in \cite{Desmars2013}, it appears that in such a case, the best solution for orbit determination is to reduce the first and  last observations rather than a random part of them. Actually, the first and last observations, by virtually increasing the length of orbital arc, bring more important constraints on the orbital motion of the asteroid. In that context, we consider five cases \citep[also considered in][]{Desmars2013} : 

\begin{itemize}
\item Case 1: we examine current observations with their current uncertainty;
\item Case 2: we assume that the first and the last observations can be reduced with the Gaia stellar catalogue;
\item Case 3: we assume that the first five and  last five observations can be reduced with the Gaia stellar catalogue;
\item Case 4: we assume that the first ten and the last ten observations can be reduced with the Gaia stellar catalogue;
\item Case 5: we assume that all observations can be reduced with the Gaia stellar catalogue.
\end{itemize}

\begin{table*}[htdp]
\begin{center}
\caption{Number and percentage of numbered NEAs for which the uncertainty of the drift will be smaller than representative values with different assumptions of new reduction with Gaia catalogue.}\label{T:Gaia}
\begin{tabular}{ccrrrrrr}
\hline
\hline
Case & Number of observations & \multicolumn{2}{c}{$\sigma_{\dot{a}}\le10^{-3}$au/My}  & \multicolumn{2}{c}{$\sigma_{\dot{a}}\le10^{-4}$au/My}  & \multicolumn{2}{c}{$\sigma_{\dot{a}}\le10^{-5}$au/My} \\
  & reduced with Gaia catalogue &  & & & & &\\
\hline
1 & 0    &    107  &  ( 6.6\%)  &   8  &   (0.5\%) &   1 &  (0.1\%) \\
2 & 2    &    346  &  (21.2\%)  &  31  &   (1.9\%) &   1 &  (0.1\%) \\     
3 & 10   &    494  &  (30.3\%)  &  93  &   (5.7\%) &   4 &  (0.2\%) \\     
4 & 20   &    644  &  (39.5\%)  & 149  &   (9.1\%) &   6 &  (0.4\%) \\     
5 & all  &   1442  &  (88.5\%)  & 688  &  (42.2\%) &  91 &  (5.6\%) \\     
\hline
\end{tabular}
\end{center}
\end{table*}

\begin{table*}[htdp]
\begin{center}
\caption{Distibution of ratio $\dot{a}_{max}/\sigma_{\dot{a}}$ with different assumptions of new reduction with Gaia catalogue.}\label{T:Gaia2}
\begin{tabular}{ccrrrrr}
\hline
\hline
Case & Number of observations &  \multicolumn{5}{c}{$\dot{a}_{max}/\sigma_{\dot{a}}$} \\
  & reduced with Gaia catalogue & [0,1] & [1,3] & [3,10] & [10,20] & $\ge20$ \\
\hline
\hline
  1 & 0   & 1567 &   47  &  14 &   0  &   1 \\
  2 & 2   & 1423 &  136  &  61 &   7  &   2 \\
  3 & 10  & 1291 &  177  & 117 &  29  &  15 \\
  4 & 20  & 1152 &  254  & 135 &  56  &  32 \\ 
  5 & all &  319 &  372  & 461 &  196 & 281 
\end{tabular}
\end{center}
\end{table*}

Table~\ref{T:Gaia} presents the distribution of drift uncertainty for the 1629 numbered NEAs according to various cases of new reduction with Gaia catalogue. 

By considering different thresholds for the uncertainty, we note that the Gaia catalogue will be helpful to get accurate detection of drift in semi-major axis. Even in the minimal case where only two observations can be reduced with the Gaia catalogue (the first and last observations), 31 NEAs will have $\sigma_{\dot{a}}\le10^{-4}$au/My, whereas only eight NEAs currently reach this level of uncertainty. The more we reduce observations with Gaia catalogue, the more the quality of the drift detection will  improve. In the extreme case where all observations could be reduced with Gaia catalogue, 688 (more than 40\% of numbered NEAs) will have $\sigma_{\dot{a}}\le10^{-4}$au/My.

Considering the maximum value of the drift, the ratio $\dot{a}_{max}/\sigma_{\dot{a}}$ also helps to measure the impact of the Gaia catalogue in the detection of drift. Table~\ref{T:Gaia2} presents the distribution of ratio for the different cases of reduction. As mentioned in previous section, the ratio can provide an approximation of the expected S/N in the case of the drift is close to its maximal value. So $\dot{a}_{max}/\sigma_{\dot{a}}\ge3$ could lead to reliable detection of the drift. Currently, only 15 objects reach this ratio, but with the minimal case with only two observations reduced with Gaia catalogue, 70 NEAs will have $\dot{a}_{max}/\sigma_{\dot{a}}\ge3$. Reducing 161 and 223 NEAs with  10 and 20 observations, respectively,  using the Gaia catalogue, will fulfill the same condition. Finally, in the extreme case where all observations could be reduced again, 938 NEAs (more than 57\%) will have a ratio greater than 3.

The Gaia catalogue will be very useful for orbit determination and for accurate detection of drift in semi-major axis, even with a few number of new reduction with Gaia catalogue. Of course, when the Gaia catalogue will be available in 2022 (or even with intermediate releases during the mission), new observations will be done and the drift uncertainty will be even better for most of NEAs. 

\section{Conclusion}
The Yarkovsky effect produces a drift in the semi-major axis of NEAs. With a precise process of orbit determination, we have detected consistent drift for 46 NEAs. The drift value we measured are in agreement with values obtained by statistical methods. Moreover, the robustness has been tested for three representative NEAs.
The drift in semi-major axis can be related to other physical parameters, such as the diameter, the obliquity, or the thermal parameter. The knowledge of the drift allows
us to constrain these physical characteristics. For example, the prograde or retrograde rotation of the NEA can be determined by the sign of $\dot{a}$. The ratio of 78\% of retrogade rotators is consistent with the expected ratio explaining the mechanism of NEA transportation.

We have also investigated the importance of precovery observations in the accurate detection of drift. In particular, we have shown that the drift uncertainty can be deeply improved by reducing only one precovery with modern techniques and modern stellar catalogue. The improvement is more important during close approach with Earth, which usually corresponds to a good period of visibility for the NEA (smaller apparent magnitude). During previous close approaches, some NEAs may have been imaged on photographic plates, and thus could provide accurate observations in order to refine orbit determination or, even more importantly, to help to detect a drift. We have already identified 14 NEAs with unused precoveries because of bad astrometry that can help to refine the detection of the drift.  

Finally, we have highlighted and quantified the impact of the Gaia catalogue in the drift determination. With this catalogue, future and also past observations could be reduced with  very high precision (a few tens of mas). If we reduce a few current observations with the Gaia catalogue, we can deeply increase the uncertainty in the drift detection. For example, with only two observations reduced again with the Gaia precision, 31 NEAs will have $\sigma_{\dot{a}}\le10^{-4}$au/My whereas only 8 NEAs currently reach this uncertainty. The Gaia catalogue opens up new perspectives in the drift detection.

The reduction of old photographic plates of solar system objects in the context of the Gaia mission is the topic of the NAROO\footnote{New Astrometric Reduction of
Old Observations} project \citep{Arlot2013}. This project  investigates natural satellites and asteroids (Transneptunian objects, NEA, or Main Belt asteroids)  in order to refine the dynamics of these objects. The detection of the drift in semi-major axis of NEAs is an important part of this project.

\begin{acknowledgements}
This work is supported by CNPq grant 161605/2012-5.
\end{acknowledgements}


\bibliographystyle{aa} 
\bibliography{biblio} 

\begin{appendix} 
\section{Maximum drift}\label{SAppendix}

To highlight unreliable detection, we introduce an estimation of a maximal drift in semi-major axis related to the absolute magnitude $H$. \cite{Vok1998} then \cite{Vok2000} gave an relation between the drift in semi-major axis due to diurnal component of the Yarkovsky effect and several physical parameters, valid for circular orbit.
We find
\begin{eqnarray}
\left(\frac{da}{dt}\right)_d=-\frac{8\alpha}{9n}\Phi(a)\frac{Gsin \delta}{1+\lambda}\cos \gamma\label{eqn1}
\end{eqnarray}

where $\alpha$ is the absortivity (complement of Bond albedo $A$), $n$ is the mean motion, $\gamma$ is the obliquity, and $\Phi(a)$ is the standard radiation function (inversely proportional to the diameter $D$, the bulk density $\rho,$ and the semi-major axis $a$). The expression  $\frac{Gsin \delta}{1+\lambda}$ is fully detailled in \cite{Vok2000} and can be approximated by 
\begin{eqnarray}
\frac{G\sin \delta }{1+\lambda} = \frac{-0.5\Theta}{1+\Theta+0.5\Theta^2}
\end{eqnarray}
where $\Theta$ is the diurnal thermal parameter.  

By using $\mu=n^2a^3$ and numerical values for $\Phi$ function, we have
\begin{eqnarray}
\left(\frac{da}{dt}\right)_d & = & 6.412\times 10^3 \frac{\alpha} {D \rho \sqrt{a}} \frac{0.5\Theta}{1+\Theta+0.5\Theta^2}\cos \gamma
,\end{eqnarray}

with the convenient units, $a$ in au and $\frac{da}{dt}$ in au/My.
The expression
$\frac{da}{dt}$ is maximum when $\gamma=0$ or 180 degrees and $\frac{0.5\Theta}{1+\Theta+0.5\Theta^2}$ reaches its maximum value 0.207 for $\Theta=\sqrt{2}$.

Finally,

\begin{eqnarray}
\left|\left(\frac{da}{dt}\right)_d\right|_{max} & = & 1.327\times 10^3 \frac{\alpha} {D \rho \sqrt{a}}
.\end{eqnarray}

To obtain a simple relation, we can consider $a=1$~au, a mean density $\rho=2600$~kg.m$^{-3}$ et a mean albedo $p_v=0.154$ for NEAs \citep{Chesley2002}. The Bond albedo will be $A=0.060$ using the relation from \cite{Bowell1989} $A=(0.290+0.684G)p_v$  (with $G=0.15$) and $\alpha=0.939$.

The diameter can be approximated by the relation \citep{Pravec2007}

\begin{eqnarray*}
D\sqrt{p_v}=1329\text{km} \times 10^{-H/5}
\end{eqnarray*}
with $H$ the absolute magnitude. 

Finally, the maximum drift can be approximated with the absolute magnitude $H$ as
\begin{eqnarray}
\left|\left(\frac{da}{dt}\right)_d\right|_{max} & = & 10^{0.2H-6.85} \text{ au/My}\label{eqmax}
.\end{eqnarray}
\end{appendix}

\end{document}